\documentstyle[12pt]{article}
\begin{document}

\title{Generalized partition functions and
interpolating statistics
\thanks{This work is supported in part by funds provided by the
U.S. Department of Energy (D.O.E.) under cooperative research
agreement \#DF-FC02-94ER40818 and Conselho Nacional de Desenvolvimento
Cient\' \i fico e Tecnol\'ogico (CNPq) - Brazilian agency.}}

\author{P. F. Borges\thanks{E-mail addresses: pborges@if.ufrj.br;
boschi@ctp.mit.edu; and farina@if.ufrj.br.},
H. Boschi-Filho$^\dagger$$^\ddagger$ and C. Farina$^\dagger$\\ 
\\ 
\it
$^\dagger$Instituto de F\'\i sica, Universidade Federal do Rio de
Janeiro \\ \it Cidade Universit\'aria, Ilha do Fund\~ao, Caixa
Postal 68528 \\ \it 21945-970 Rio de Janeiro, BRAZIL\\ \it and \\ 
\it
$^\ddagger$ Center for Theoretical Physics\\ \it Laboratory for
Nuclear Science\\ \it Massachusetts Institute of Technology\\ \it
Cambridge, Massachusetts 02139-4307, USA}

\bigskip
\maketitle
\begin{abstract}
We show that the assumption of quasiperiodic boundary conditions
(those that interpolate continuously periodic and antiperiodic
conditions) in order to compute partition functions of relativistic
particles in 2+1 space-time can be related with anyonic physics. In
particular, in the low temperature limit, our result leads to the well
known second virial coefficient for anyons.  Besides, we also obtain
the high temperature limit as well as the full temperature dependence
of this coefficient.
\end{abstract}

\bigskip
\bigskip

\pagebreak

In nature one encounters particles which obey Bose-Einstein (BE) or
Fermi-Dirac (FD) statistics. However, speculations concerning the
existence of particles which obey intermediate (fractional) statistics
between BE and FD have been raised for a long time. In recent years
this kind of conjecture has gained force with its possible relation
with condensed matter effects like high-$T_C$ superconductivity and
the fractional quantum Hall effect \cite{review}.

In $(2+1)$ dimensional field theory, fractional statistics can be
obtained by including a Chern-Simons term \cite{DeserJT} in the
Lagrangian density describing the interaction of matter with gauge
fields.  Particles with intermediate statistics between the BE and FD
cases in two space dimensions are usually called anyons.  Their
properties can be understood from the braid group or equivalently
assuming that they are hard core indistinguishable particles in two
space dimensions so that the configuration space becomes a manifold
$\cal M$ with a non-trivial topology in the sense that $\Pi_1 ({\cal
M}) =B_N$ for $N$ particles, where $B_N$ denotes the braid group with
$N$ generators.  Hence, it can be shown that when anyons turn around
each other the corresponding wave function (or the Feynman propagator
for the system) acquires a non-trivial geometric phase \cite{Leinaas
and Myrhein}--\cite{Lerda}. An inherent difficulty of the anyon field
description is its non-locality \cite{Frolich}.  This makes it
difficult to calculate beyond the second virial coefficient
\cite{Arovas} even in the case of ``free'' anyons, contrary to what
happens for free bosonic and fermionic gases for which the full
partition function is known. 

Here, we shall discuss a different approach to the interpolation
between bosons and fermions. In this paper we shall compute a
generalized partition function which contains the relativistic
partition functions for bosons and fermions as particular cases, and
also interpolates continuously between these cases. From this
partition function we obtain the second virial coefficient and show
that, in the low energy (temperature) limit it reproduces correctly
the results known for anyons in the literature \cite{Arovas}. We also
find the high temperature and the full temperature dependence of the
relativistic second virial coefficient. 

Let us briefly review how to obtain the partition function for ideal
relativistic bosonic and fermionic gases, from which we will
generalize to the anyonic case. The partition function for a system
described by a Hamiltonian $H$ with chemical potential $\mu$ can be
written as
\begin{equation}\label{Z}
{\cal Z}\equiv \exp\{-\beta\Omega\}= Tr \;e^{-\beta(H+ {\cal N}\mu)}, 
\end{equation}

\noindent where $\Omega$ is the free energy and $\cal N$ is the
conserved charge of the system. As it is well known for relativistic
charged massive bosonic fields one can express this partition function
as a determinant \cite{Bernard}\cite{Kapusta}, namely,
\begin{equation}\label{bosondet}
{\cal Z}=\left[\left.\det(-D^2+M^2)\right|_{P}\right]^{-1}
;\;\;\;\;\;\;\;\;(bosons)\;, 
\end{equation}

\noindent where $D^2$ is the square of the covariant derivative,
$D_\nu$, including the chemical potential, $D_\nu=(\partial_0 +
i\mu,\partial_i)$ and $M$ is the mass of the field.  This prescription
for the inclusion of the chemical potential $\mu$ coincides with the
one given by eq. (\ref{Z}) except for the introduction of an imaginary
part to $\cal Z$, which is not physically relevant.  So unless
otherwise specified, we are considering that the real part of $\cal Z$
has been taken. Note also that, the chemical potential appears squared
in the relevant operator $-D^2+M^2$.  This is a consequence of
representing the partition function in phase space and then
integrating over momenta (see \cite{Kapusta} for details).  The label
{\it P} means that the eigenvalues of this operator are subjected to
periodic boundary conditions and hence are given by
\begin{equation}\label{eigenvalues}
\lambda_{nk}=(\omega_n + i\mu)^2+\vec k^2+M^2\;, 
\end{equation}

\noindent where $\omega_n=2n\pi/\beta$, with $n\in {{\sf Z}\!\!{\sf
Z}}$, are the Matsubara frequencies \cite{Matsubara} for bosonic fields
and $\vec k\, \in \, {{\sf I}\! {\sf R}}^N$.

As for the case of bosons, one can also write the partition function
for free relativistic fermions as a determinant
\begin{eqnarray}\label{fermiondet} {\cal Z} & = &
\left.{\det}_D(i\slash\!\!\!\!D -M)\right|_{A}
\nonumber\\ & = & \left[\left.\det(-D^2+M^2)\right|_{A}\right]^{d/2}
;\;\;\;\;\;\;\;\;(fermions), 
\end{eqnarray}

\noindent where the determinants are calculated over the corresponding
operator and $\det_D$ means the inclusion of the calculation over
Dirac indices ($d$ is the dimension of the Dirac representation which
hereafter we take $d=2$). The subscript {\it A} means that the
eigenvalues of $-D^2+M^2$ are now computed with antiperiodic boundary
conditions.  Hence, they are given by (\ref{eigenvalues}), but now the
Matsubara frequencies are $\omega_n=(2\pi/\beta)(n+1/2)$, due to the
antiperiodic boundary condition.

The remarkable fact about the (last line of) equation
(\ref{fermiondet}) is the well known but few explored character that
it states that it is possible to compute the partition function for
fermions through an equation where spin does not appear at all. The
information that the partition function (\ref{fermiondet}) corresponds
to fermions is solely described by the fact that the determinant of
the Klein-Gordon operator $(-D^2+M^2)$ is computed over antiperiodic
eigenfunctions. Naturally, this antiperiodicity comes from the fact
that fermions anticommute when they are exchanged in space, {\sl
i. e.}, the wave function of the system acquires a sign $(-1)$. So, on
thermodynamical grounds the only necessary information to compute a
partition function for free fermions is that they obey energy
conservation (expressed through the Klein-Gordon operator) and that
their wave functions are antiperiodic in the Euclidean time.

Now, we want to extend this fundamental result to the case of
(quasi) particles that obey an intermediate statistics between the
bosonic and fermionic cases. As is well known these particles live in
2+1 dimensional space-time and they can be described by specific 
equations of motion like the one of Jackiw and Nair
\cite{Jackiw&Nair}. This equation of motion governs the dynamics
of that particles and contains detailed information on their algebraic
properties, including the braid group as well. However, the task of
computing a determinant of an operator obtained from such equation is
very involved and we are not going in this direction in this work.

An alternate route, that we are going to follow here, is to use very
few but essential information which are necessary for a
thermodynamical description of a system of such particles. These
informations are only three: first these (relativistic) particles
should obey energy conservation and so in spite that they are
described by some involved specific 2+1 dimensional equation they
should also obey the Klein-Gordon equation; second the determinant of
the Klein-Gordon operator should be calculated using some appropriate
boundary condition; Finally the partition function is defined by a
certain power $\sigma$ (to be discussed later) of the determinant of
the Klein-Gordon operator, under a chosen boundary condition.  Of
course, if the boundary condition is periodic in Euclidean time and
$\sigma=-1$ we will be describing bosons as is expressed in equation
(\ref{bosondet}). On the other hand, if we impose antiperiodic
boundary conditions and $\sigma=+1$ we will be describing fermions as
in equation (\ref{fermiondet}). As we want to describe particles that
obey some statistics that continuously interpolate the bosonic and
fermionic cases, it seems rather natural to choose a boundary
condition that interpolates these two well known cases. Such a
condition, which we call quasiperiodic, can be written as
\begin{equation}\label{condition}
\psi(0;\vec x)=e^{i\theta}\psi(\beta ;\vec x). 
\end{equation}

\noindent 
where $\theta$ is a continuous parameter defined in the interval
$[0,\pi]$. Note that putting $\theta=0$ we reduce the above condition
to a periodic boundary condition and if $\theta=\pi$ it becomes an
antiperiodic one.

The power $\sigma$ of the determinant is introduced to fit the bosonic
and fermionic cases. As is well known in these cases the powers $-1$
and $+1$, respectively, come from the definition of the measure over
c-number (bosons) and Grasman variables (fermions). For the anyonic
case it would be necessary to define a set of variables which could
interpolate the properties of c-numbers and Grasman variables. This
approach seems to be related to q-deformed calculations, but here we
are just going to leave the power $\sigma$ of the determinant free and
check, at the end of the calculations that both $\sigma=\pm 1$ will be
related to anyonic physics. This may be explained remembering that
conventional anyons are constructed from fermions or bosons (with a
fixed built in measure) adding a Chern-Simons term. Later on, we will
also show that keeping $\theta$ fixed and varying $\sigma$ we could
describe as well the interpolation between bosons and fermions.

Note that the quasiperiodic condition (\ref{condition}) can also
follow from an analogy of the spatial behavior of bosons, fermions and
anyons. This behavior is the well known property that bosonic wave
functions are symmetrical while fermionic ones are antisymmetrical and
anyonic wave functions acquire a phase $\exp(i\theta)$ when these
particles are turned around each other \cite{review,Jackiw90}.  Going
to the finite temperature case, these properties imply that bosonic
fields are periodic in Euclidean time while fermionic are
antiperiodic. This analogy is not complete for anyons and we are
assuming here that an intermediate continuous condition can explain at
least some of the thermodynamical properties of anyons, as we are
going to show bellow.

Then, we are not proving that anyons obey the quasiperiodic boundary
condition, eq. (\ref{condition}), and we are not able at this point to
do so, but we are using a very appealing analogy to support it. The
condition (\ref{condition}) is our basic assumption from which we are
going to show that the partition function that we will calculate can
be mapped onto the corresponding results for conventional anyons. In
fact we do this for the only exact result known for anyons in this
matter which is the second virial coefficient obtained by Arovas {\sl
et. al.} \cite{Arovas}. The calculation of higher virial coefficients
and its comparison with the perturbative or numerical ones known in
the literature is done in a forthcoming work \cite{BBF}.

The price we have to pay in our approach is that we are loosing
detailed quantum mechanical information as those associated with the
braid group which become very involved when many particles come into
play. The obvious advantage of this global approach is that it keeps
only the essential thermodynamical information which suffices for
calculating the virial coefficients, for example.

Note also that, in our approach there is no reason for restricting the
space-time dimensions, despite that anyons live only in two space
dimensions. This is not surprising since other approaches towards
generalized statistics, like the Haldane's generalized exclusion
principle \cite{Haldane}, can be formulated in spaces with arbitrary
number of dimensions. In fact, at the end of our calculations we shall
restrict them to two space dimensions in order to show the equivalence
of our approach with the conventional one for anyons.

Let us now compute the following determinant of the Klein-Gordon
operator:
\begin{equation}\label{det}
{\cal Z}=\left[\left.\det(-D^2+M^2)\right|_{\theta}\right]^{\sigma}, 
\end{equation}

\noindent where the subscript $\theta$ means that the quasiperiodic
condition (\ref{condition}) is assumed and we introduced the parameter
$\sigma$ to be able to reproduce correctly the bosonic and fermionic
particular cases. Observe that when $\theta=0$ and $\sigma=-1$ we
reobtain the bosonic partition function, while for $\theta=\pi$ and
$\sigma=+1$ we have the fermionic case. For generality, we are going
to calculate the above determinant in $N+1$ space-time dimensions.
The eigenvalues of the Klein-Gordon operator in this case are: 
\begin{equation}\label{eigenvalues2}
\lambda_{n k\theta}= \left[{2n\pi \over \beta} +i\mu + {\theta\over
\beta}\right]^2 +\vec k^2+M^2\;. 
\end {equation}

\noindent The determinant (\ref{det}) is a generalization of its
quantum mechanical $(0+1)$ dimensional version (with $\mu=0$) which
has been calculated using Green functions \cite{BFD} and the zeta
function method \cite{BF}. Here, we are going to calculate this
determinant also using the generalized zeta function method through
the basic formula \cite{Hawking} 
\begin{equation}\label{zeta}
\det{\cal A} =\left.\exp\left\{- {\partial \over \partial
s}\zeta(s;{\cal A}) \right\} \right|_{s=0}, 
\end{equation}

\noindent where the generalized zeta function is defined by
$\zeta(s;A)=Tr\,A^{-s}$ and an analytical continuation of $\zeta(s;A)$
to the whole complex plane of $s$ is tacitly assumed. 

For the case at hand, with ${\cal A}=-D^2+M^2$ under $\theta$-periodic
boundary conditions, the eingenvalues being given by
eq. (\ref{eigenvalues2}), the generalized zeta function reads
\begin{equation}
\zeta(s;{\cal A})={V a_N \over (2\pi)^{N}}
\sum_{n=-\infty}^{+\infty}\int_0^\infty dk\ k^{N-1} \left(
\left[{2n\pi \over \beta} +i\mu + {\theta\over \beta}\right]^2 +\vec
k^2+M^2 \right)^{-s}\;, 
\end{equation}

\noindent where $V$ is the volume, $a_N$ is the area of the unit
hypersphere, both in $N$ space dimensions. The integral is expressible
in terms of the Beta function, and we obtain
\begin{equation}\label{zeta(s,A)}
\zeta(s;{\cal A})={C_N \over 2}{\Gamma({N \over 2}) \Gamma(s-{N
\over2}) \over \Gamma(s)}({2\pi \over \beta})^{-2s+N}
\sum_{n=-\infty}^{+\infty} [\nu^2+(n+\chi)^2]^{-(s-{N \over 2})}\;, 
\end{equation}

\noindent \noindent where we have defined 
\begin{equation}\label{defs}
C_N={V a_N \over (2\pi)^N}\;;\;\;\;\;\; \nu={\beta M\over 2\pi}\;;
\;\;\;\;\;\chi={i\beta\mu+\theta \over 2\pi}\;. 
\end{equation}

\noindent The sum appearing in eq. (\ref{zeta(s,A)}), well defined for
$\Re (s)>1+{N\over 2}$, is a generalization of the usual Epstein
function \cite{Epstein} but as one can see from formula (\ref{zeta}),
that the region of interest contains the point $s=0$. The analytic
continuation of this sum for the whole complex s-plane can be written
as \cite{Kirsten1}:
\begin{eqnarray}
&& \sum_{n=-\infty}^{+\infty}[\nu^2+(n+\chi)^2]^{-(s-{N \over 2})}
\nonumber \\ &=&{\sqrt\pi \over\Gamma(s-{N\over 2})}
\left[{\Gamma(s-{N\over 2}-{1\over
2})\over\nu^{2s-N-1}}+4\sum_{n=1}^{+\infty}cos(2\pi n\chi).({n\pi
\over \nu})^{s-{N\over 2}-{1\over 2}} K_{s-{N\over 2}-{1\over 2}}(2\pi
n \nu)\right]\;, 
\end{eqnarray}

\noindent where $K_{\alpha}(z)$ is the modified Bessel function of
order $\alpha$. Taking the derivative of $\zeta(s,\cal A)$ with
respect to $s$, using that $lim_{s\rightarrow 0}[1/\Gamma(s)]=0$ and
$lim_{s\rightarrow 0}[(d/ds)\Gamma(s)]=1$ we have, apart from an
irrelevant term which is linear in $\beta$ and independent of $\mu$: 
\begin{equation}
\left.  {\partial\over\partial s}\zeta(s,{\cal A})\right|_{s=0} =
2{\sqrt\pi}C_N \Gamma({N \over 2}){\left({2\pi \over
\beta}\right)}^{N} \sum_{n=1}^{+\infty}\cos(2\pi n
\chi){\left({n\pi\over \nu}\right)}^{-{1 \over 2}(N+1)} K_{-{1 \over
2}(N+1)} (2\pi n \nu)\;. 
\end{equation}

\noindent So, using the fact that $\ln \det {\cal A}= -
(\partial/\partial s) \zeta (s,{\cal A})\vert_{s=0}$, we can find the
generalized determinant (\ref{det}) which interpolates continuously
the partition functions, or equivalently the free energy,
eq. (\ref{Z}), for relativistic boson and fermion gases with chemical
potential $\mu$ in $(N+1)$-dimensions: 
\begin{eqnarray} 
\Omega(\beta,\mu)&\equiv& - {1\over \beta} \ln {\cal Z} 
= - {1\over \beta} \sigma \ln \det {\cal A}\nonumber\\
&=&\sigma {V a_N \over (2\pi)^N} {1\over \sqrt\pi}
\left({ 2M\over \beta}\right)^{{1 \over 2}(N+1)} \Gamma({N \over 2})
\nonumber \\ &\times&
\sum_{n=1}^{+\infty}\cos(n\theta)\cosh(n\beta\mu) ({1 \over n})^{{1
\over 2}(N+1)}K_{{1 \over 2} (N+1)}(n\beta M)\;, 
\label{Omega}
\end{eqnarray}

\noindent where $C_N$, $\nu$ and $\chi$ have been taken from
eq. (\ref{defs}).  Note that in the above formula, only the real part
of the free energy $\Omega(\beta,\mu)$ was taken into account,
according to the prescription of introducing the chemical potential as
an imaginary time-component gauge potential \cite{Kapusta}. If we
particularize the parameters $\sigma$ and $\theta$ to the bosonic
($\sigma=-1$ and $\theta=0$) and fermionic ($\sigma=+1$ and
$\theta=\pi$) cases we shall find precisely the results known in the
literature \cite{HW}-\cite{Actor}.

To show the equivalence of this partition function with the
conventional theory of nonrelativistic anyons, we first recall the
cluster expansion
\begin{equation}
{1\over V} \ln {\cal Z} = \sum_{l}\; b_l(V,T)\; z^l\;,
\end{equation}

\noindent where $z\equiv\exp{\beta\mu}$ is the fugacity and $b_l(V,T)$
are the usually called cluster coefficients. As ${\cal Z}=\exp (-\beta
\Omega)$, we substitute the expression of $\Omega(\beta,\mu)$,
eq. (\ref{Omega}), into the above equation and find that the cluster
coefficients, in our case, are given by
\begin{equation}
b_{\pm n} (V,T) 
= - \; 2\sigma \; {\left({M\over 2\pi n}\right)}^{N+1\over 2}
\beta^{1-N\over2}  \cos(n\theta)\; K_{N+1\over 2} (n\beta M)\;;
\;\;\;\;(n=1,2,3,...)\;.
\end{equation}

\noindent The $\pm$ sign for the cluster coefficients are related to
particles and antiparticles. Note that, in a conventional
nonrelativistic system only particles (or antiparticles) are
present, so in that case we would have found only one set of natural
cluster coefficients. As they are in fact equal, as required by
particle-antiparticle symmetry, this will not modify our analysis.

Now, using the standard formula which relates the second virial
coefficient with the cluster ones \cite{Dash},
\begin{equation}
B(T)= -\; {b_2(V,T)\over [b_1(V,T)]^2}\;,
\end{equation}

\noindent
we have in $N$ space dimensions the exact expression for the
relativistic second virial coefficient interpolating the bosonic and
fermionic cases:
\begin{equation}\label{BN}
B(T) = {1\over 2\sigma}\; {\left({\pi \over M }\right)}^{N+1\over 2}
\beta^{N-1\over2} \; (1-\tan^2\theta)\; { K_{N+1\over 2} (2 \beta M)
\over {\left[ K_{N+1\over 2} (\beta M)\right]}^2}\;.
\end{equation}

To write the ratio of the two Bessel functions in a more familiar form
we can use the asymptotic expression for them
\begin{equation}\label{asymp}
K_{\nu} (x) = \sqrt{\pi \over 2x} \; e^{-x}\;
\left[1+{4\nu^2-1\over 8x}+ {(4\nu^2-1)(4\nu^2-3^2)\over 2!\;
(8x)^2}+\dots\right]\;,
\end{equation}

\noindent valid when $-\pi/2 < \arg x < \pi/2$. Once we are interested
in relating our relativistic partition function to the conventional
formulation of nonrelativistic anyons, we must reduce our formula by
taking the low temperature (nonrelativistic) limit $\beta M>>1$ on
equation (\ref{BN}) of the second virial coefficient. In this case,
the above asymptotic expression for the Bessel functions can be
greatly simplified to
\begin{equation}\label{KL}
K_{\nu} (x) \simeq \sqrt{\pi \over 2x} \; e^{-x}\;;
\;\;\;\;\;\;\;\;\;\;\;\;
(x>>1)\;,
\end{equation}

Form now on, we will restrict our analysis to the particular case of
interest, which is $N=2$ space dimensions where anyons live. So
considering the low temperature approximation and using
eq. (\ref{KL}), we find that eq. (\ref{BN}) reduces to $(\hbar=c=1)$:
\begin{equation}\label{BL}
B(T) = {1\over 2\sigma} {\pi \beta \over M }\; (1-\tan^2\theta)\;;
\;\;\;\;\;\;\;(\beta M >> 1).
\end{equation}

\noindent We show now that this is equivalent to the results of Arovas
{\sl et. al.} for the second virial coefficient. If one starts from
charged fermions interacting with a magnetic flux tube with intensity
$\phi=\alpha h c /e$, one gets \cite{Arovas}
\begin{equation}\label{af}
B(T) = {1\over 4}\; \lambda_T^2\; (1-2\delta^2)\;,
\end{equation}

\noindent or analogously, starting from bosons the result is 
\begin{equation}\label{ab}
B(T) = - {1\over 4}\; \lambda_T^2\; (1-4|\delta|+2\delta^2)\;,
\end{equation}

\noindent where $\lambda_T^2={2\pi\hbar^2\over MkT}$ is the thermal
wavelength and $\delta$ is the non-integer part of the statistical
parameter $\alpha$, {\sl i. e.}, $\alpha=2j+1+\delta$ for fermions,
eq. (\ref{af}), and $\alpha=2j+\delta$ for bosons, eq. (\ref{ab}),
$j=0,\pm 1,\pm 2,...$ in both cases.

To compare these results, we note that, in our case, if we start from
fermions we should put $\sigma=+1$, identify $\tan^2\theta=2\delta^2$
and get the same answer for the second virial coefficient of anyons,
as in eq. (\ref{af}). Note that these fermions came about through the
choice $\theta=\pi$, but we can redefine $\theta=\theta^\prime +n\pi$
such that $-\pi/2\le \theta^\prime\le +\pi/2$, since the $\theta$
dependence appears only in the second virial coefficient through
$\tan^2\theta$. In fact, if we want to {\sl interpolate} the virial
coefficients of bosons $(-\lambda_T^2/4)$ and fermions
$(+\lambda_T^2/4)$ we must restrict $\tan^2 \theta \le 2$. This
restriction, which can be written as $- \arctan \sqrt{2}\le
\theta^\prime \le \arctan \sqrt{2}$, is exactly the one which is
needed to map our second virial coefficient onto the one of Arovas
{\sl et. al.}

In the other case, if we start with bosons, we put $\sigma=-1$ and
identify $\tan^2\theta=2|\delta|(2-|\delta|)$, in a complete analogous
situation where the same restrictions on $\theta$, from the fermionic
case, applies here as well.  One should also note that, the Arovas
{\sl et. al.}  second virial coefficient presents cusp singularities
for $\Delta \alpha =2$, which are not present in our result since the
coefficient we found is analytical in its parameters except for poles
at $\theta=(n+1/2)\pi$, where $n=0,\pm 1, \pm 2, ...$ This difference
comes from the fact that our result also contains values for $B(T)$
which extrapolate the interval $(-\lambda_T^2/4,+\lambda_T^2/4)$.

A simpler situation comes up if we consider the particular cases of
small angles $\theta\simeq 0$ or $\theta\simeq \pi$, so that
$$\tan^2\theta\simeq \theta^2.$$

\noindent
In these situations the relation between the quasiperiodic $\theta$
and the statistical parameter $\alpha$ (or $\delta$) becomes simply
(considering the fermionic anyon model given by (\ref{af})):
$$\sqrt{2}\alpha = \theta.$$

So, in the region where anyons are close to fermions (or bosons),
where usually perturbation calculations are done for other quantities
like the higher virial coefficients, $\theta$ plays the role of the
statistical parameter $\alpha$ up to a constant factor $\sqrt{2}$. Of
course, the exact relation between these parameters is nonlinear as
found above, but in that case the analysis is much more involved and
we do not have until now a deeper understanding of this nonlinear
behavior.

To understand how we succeed in describing the statistical mechanics
of anyons in this simple way, we may say that, the interpolating
parameter $\theta$ discussed here plays the role of a topological
constant gauge field $A_\mu=(A_0,\vec 0)$, since the $\theta$-boundary
condition for the Klein-Gordon operator, both in the bosonic as well
as in the fermionic case, implied that the Matsubara frequencies
change according to $\omega_n\rightarrow \omega_n +\theta/\beta$,
which can be viewed as a shift in the time derivative operator,
$\partial_0\rightarrow\partial_0 +i\theta/\beta$. As the time
coordinate, $x_0$, is compactified to the interval $(0,\beta)$, this
introduces the non-trivial topology usually associated with anyons but
in a different and unexpected manner. This prescription, which works
{\sl only} for the finite temperature case, allows us to relate
directly the quasiperiodic boundary condition with intermediate
statistics, or anyon description.

Another interesting point in our formulation is that we can
interpolate the bosonic and fermionic second virial coefficients, from
two different points of view. First, keeping $\sigma$ fixed to $\pm 1$
and varying $\theta$ we were able to describe this interpolation very
close to the anyon description starting from bosons and fermions with
their original equations of motion, or equivalently the bosonic and
fermionic determinants, with the inclusion of a topological term, in
our case the $\theta$-boundary condition, as we described
above. Alternatively, we can keep $\theta$ fixed to $0$ or $\pi$ and
vary continuously $\sigma$ between $+1$ and $-1$, so that we should
identify $\sigma^{-1}=1-2\delta^2$ in the fermionic case and
$\sigma^{-1}=1- 2|\delta|(2-|\delta|)$ in the bosonic case. This seems
to be more close to a q-deformed calculation but this connection is,
at least for the moment, far from being trivial.

We can go even further in the determination of the relativistic anyon
second virial coefficient, since we do not need to use the low
temperature approximation in our calculations. This is so because, in
our exact result, eq. (\ref{BN}), we can also use (in $N=2$) the exact
expression for the Bessel function of order 3/2:
\begin{equation}
K_{3\over 2} (x) = \sqrt{\pi \over 2x}\; e^{-x} \left( 1+{1\over x}
\right)\;.
\end{equation}

Substituting this expression in the second virial coefficient,
eq. (\ref{BN}), instead of the approximate one, eq. (\ref{KL}), we
find the exact relativistic result for the anyon second virial
coefficient:
\begin{equation}\label{N2BT}
B(T) = {1\over 2\sigma} {\pi \beta \over M } (1-\tan^2\theta)
\left(1+{1\over 2\beta M}\right) {\left(1+{1\over \beta
M}\right)}^{-2}. 
\end{equation}

\noindent
Note that the possibility of finding this exact result for the second
virial coefficient is a peculiarity of odd-dimensional space-times,
since for these cases the relevant Bessel functions are of
half-integer order for which the asymptotic expansion became exact
(see eq. (\ref{asymp})). Naturally, the low temperature result,
eq. (\ref{BL}), can be easily obtained from (\ref{N2BT}) just taking
the low temperature limit $\beta M >> 1$.

Furthermore, we can also find the high temperature limit (extremely
relativistic case) of expression (\ref{N2BT}) taking $\beta M <<1$ so
that we find
\begin{equation}
B(T) = {1\over 4\sigma}\; \pi \beta^2 \; (1-\tan^2\theta)
\;;\;\;\;\;\;\;\;\;(\beta M <<1)\;.
\end{equation}

Note that the statistical correction $(1-\tan^2\theta)$ is independent
of the high/low temperature regime, but the mass dependence disappears
in the high temperature case, while the power of temperature just
changes from $\beta$ to $\beta^2$.

As a final comment let us mention that it seems interesting to
investigate the extension of these results to other related
situations. One of these is the calculation of the higher virial
coefficients which are very promising since they can be exactly
evaluated in our approach \cite{BBF}. Another situation of interest is
the inclusion of an external magnetic field since in most applications
of anyonic physics, like the fractional quantum Hall effect, strong
external fields are present. We should report on this elsewhere.

\bigskip
\bigskip

\noindent {\bf Acknowledgments.} 
We thank M. Plyushchay for interesting discussions on relativistic 2+1
dimensional equations of motion, K.  Kirsten for calling our attention
to refs.  \cite{Kirsten1}, where appear the analytic continuation we
needed for the calculation of our generalized zeta function and
G. A. Goldin for interesting comments on the manuscript and for
calling our attention to refs. \cite{GMS}. H.B.-F.  acknowledges
Professor R. Jackiw for calling our attention that nonrelativistic
anyons at zero temperature obey a {\sl spatial} quasiperiodic boundary
condition \cite{Jackiw90}, for a careful reading of the manuscript,
many useful suggestions and his hospitality at the Center for
Theoretical Physics - MIT.  The authors H.B.-F. and C.F. were
partially supported by CNPq (Brazilian agency).

\end{document}